\documentclass[pra,showpacs,floatfix]{revtex4}

\usepackage{graphics,graphicx}
\usepackage{mathrsfs}
\usepackage{amsmath,amsfonts,amssymb}
\usepackage{subfigure}

\newcommand{\f}{\frac}
\newcommand{\pa}{\partial}

\begin{document}

\title{Far-off-resonant wave interaction in one-dimensional photonic crystals
       with quadratic nonlinearity}

\author{Johannes-Geert Hagmann 
        \footnote{Present address: Laboratoire de Physique, 
                                   Ecole Normale Sup\'erieure de Lyon, 
                                   69364 Lyon, France}
       }
\affiliation{Institut f\"ur Theoretische Festk\"orperphysik, 
             Universit\"at Karlsruhe (TH), 76128 Karlsruhe, Germany}

\author{Lasha Tkeshelashvili}
\affiliation{Institut f\"ur Theoretische Festk\"orperphysik, 
             Universit\"at Karlsruhe (TH), 76128 Karlsruhe, Germany}

\author{Guido Schneider}
\affiliation{Institut f\"ur Analysis, Dynamik und Modellierung, 
             Universit\"at Stuttgart, Pfaffenwaldring 57, 
              70569 Stuttgart, Germany}

\author{Kurt Busch}
\affiliation{Institut f\"ur Theoretische Festk\"orperphysik, 
             Universit\"at Karlsruhe (TH), 76128 Karlsruhe, Germany}

\date{\today}

\begin{abstract}
\noindent

We extend a recently developed Hamiltonian formalism for nonlinear
wave interaction processes in spatially periodic dielectric structures 
to the far-off-resonant regime, and investigate numerically the three-wave 
resonance conditions in a one-dimensional optical medium with 
$\chi^{(2)}$ nonlinearity.  
In particular, we demonstrate that the cascading of nonresonant 
wave interaction processes generates an effective $\chi^{(3)}$ 
nonlinear response in these systems. We obtain the corresponding 
coupling coefficients through appropriate normal form transformations 
that formally lead to the Zakharov equation for spatially periodic 
optical media. 

\end{abstract}


\maketitle

\section{Introduction \label{int}}
During the last two decades, the emergence of periodically 
structured optical materials -- commonly called photonic 
crystals (PCs) -- has lead to substantial progress in the 
science and technology of optics and photonics. For a recent
review, we refer to Ref. 
\cite{PhysRep}. 
It has soon been realized that PCs with nonlinear constituent
materials facilitate the realization of novel effects and
functionalities that are very hard to realize with other 
systems. 
However, a mathematically consistent description of wave-interaction 
processes in PCs is complicated due to a number of formal and as 
yet unsolved problems relating to the complex structure of the 
equations of motion. Recently, based on a Hamiltonian formulation
of Maxwell's equations, an effective field theory for nonlinear 
PCs has been developed by Volkov and Sipe
\cite{vs}. 
This approach allows a systematic study of nonlinear 
effects beyond the commonly used multiple-scales analysis
\cite{degm}. 
In particular, in Ref. \cite{vs}, the formalism has been applied 
to the problem of resonant interactions of wave packets in PCs
with quadratic nonlinear response. The purpose of this article 
is to extend this formalism to the far-off-resonant, i.e. nonresonant, case in 
one-dimensional systems. 
In particular, we reduce the nonlinear equations of motions of
corresponding nonlinear PCs to the simplest form possible. This 
reduction is facilitated through the systematic elimination of 
nonresonant interaction terms from the classical Hamiltonian
via appropriate normal form transformations. This procedure
results in the emergence of an effective third-order response
of the system. We would like to note that similar techniques 
have been successfully employed in the analysis for magnetic
systems 
\cite{lvov}
as well as for ideal fluid hydrodynamic waves
\cite{zakharov}. 
In the latter case, the result is the so-called Zakharov equation 
for surface water waves, 
which has recently been studied experimentally 
\cite{skj}. 

However, in contrast to the case of spin or water waves, where 
the underlying linear dispersion relations prohibit nontrivial 
three-wave resonances, such resonances do appear in PCs for certain
sets of interacting wave numbers. 
This leads to nonremovable singular terms in the normalized 
Hamiltonian. Consequently, additional constraints on the 
wave packets are required in order to mathematically justify the 
validity of the simplified equations for finite times. 
In realistic physical systems, where losses due to fabricational 
tolerances and/or residual material absorption, or simply the finite 
sample size provide natural upper bounds on the time scales, these
constraints may be less severe. 

The article is organized as follows. 
In Sec. \ref{ref1}, we extend the Hamiltonian formalism of 
Volkov and Sipe \cite{vs} for 
one-dimensional photonic crystals with $\chi^{(2)}$ nonlinearity
to the nonresonant case.
In Sec. \ref{ref2}, we analyze numerically the lowest-order
resonance conditions and compare the results to the water wave 
problem for which the spectrum is known analytically. In Sec. 
\ref{ref3}, the normal form reduction of the Hamiltonian is
performed and the Zakharov equation for nonlinear PCs is formally
derived for the case when three-wave interaction processes are
suppressed. 
In particular, the expressions for the effective third-order 
nonlinear response are derived and compared with the case of
homogeneous dispersive $\chi^{(2)}$ materials. We conclude with a 
discussion of the limitations of our approach, and provide
 an outlook on possible future work in Sec. \ref{ref4}.

\section{Classical Hamiltonian formulation \label{ref1}}
In this section we derive a classical Hamiltonian formulation for 
nonlinear PCs. The derivation closely follows Ref. 
\cite{vs}, however, we emphasize that, as compared to Ref.
\cite{vs}, 
we have chosen a different set of fundamental fields
for the decomposition into a complete set of eigenfunctions.

We consider infinite nonlinear PCs
with nonmagnetic, dispersion- and lossless constituent materials. 
Since the linear dispersion provided by photonic crystals' microstructure
is typically much larger than the material dispersion of its constituent
materials, this is a very good approximation. If material dispersion
dominates -- such as would be the case for quasiphase matched
gratings where only the nonlinear properties vary periodically in space -- 
one would have to resort to a slowly varying envelope approximation
\cite{cbk}.
Assuming further that the material response is local in space and 
time, we can express the polarization $\mathbf{P}(\mathbf{x},t)$ 
that connects the dielectric displacement $\mathbf{D}(\mathbf{x},t)
   =   \epsilon_0 \mathbf{E}(\mathbf{x},t)+ 
                  \mathbf{P}(\mathbf{x},t)$ 
with the electric field $\mathbf{E}(\mathbf{x},t)$ as 
\begin{eqnarray}
D_i(\mathbf{x},t)
  & = & \epsilon_0 E_i(\mathbf{x},t)
          + \chi^{(1)}_{i j}(\mathbf{x})
            E_{j}(\mathbf{x},t) 
          +\chi^{(2)}_{ijk}(\mathbf{x})
            E_{j}(\mathbf{x},t)E_k(\mathbf{x},t)
+ \dots\ . \label{response}
\end{eqnarray}
Here, $\epsilon_0$ denotes the free-space dielectric permittivity.
The susceptibility tensors $\chi^{(n)}$ of rank $n+1$ are 
symmetric under permutations of their indices and -- here as 
well as in the remainder of the article -- summation over
repeated indices is implied. Instead of the first-order
susceptibility $\chi^{(1)}$, we employ the dielectric tensor 
$\epsilon(\mathbf{x})$. In addition, in the following, we will
restrict ourselves to the case of isotropic linear response, so
that we have 
$\chi^{(1)}(\mathbf{x}) = 
 \epsilon_0 \left(\epsilon(\mathbf{x})-1 \right)$.

Throughout this paper, we identify the canonical momentum
with the displacement field $\mathbf{D}$ rather than with 
the electric field $\mathbf{E}$. Therefore, it is desirable 
to expand the electric field in terms of the displacement field 
according to
\begin{eqnarray}
E_i(\mathbf{x},t)
  = \epsilon_0^{-1} \epsilon(\mathbf{x})^{-1} D_i(\mathbf{x},t)
     + \Gamma_{ijk}^{(2)}(\mathbf{x})
        D_j(\mathbf{x},t) D_k(\mathbf{x},t)+\dots\ .
\end{eqnarray}
For the second-order susceptibility, the relations between
$\chi^{(2)}$ and $\Gamma^{(2)}$ reads
\begin{eqnarray}
\Gamma^{(2)}_{ijk}(\mathbf{x})
   & = & 
  - \frac{\chi^{(2)}_{ijk}(\mathbf{x})}{(\epsilon_0\epsilon(\mathbf{x}))^{3}},
\end{eqnarray}
and the relations for the higher-order susceptibilities can be 
found by recursion \cite{hm}. In PCs, the material properties are
periodic functions with respect to the set of lattice vectors 
$\mathbf{R}$, i.e., we have 
$ \epsilon(\mathbf{R}+\mathbf{x})
  = \epsilon(\mathbf{x})$,
$\chi_{i_1i_2...i_n}^{(n-1)}(\mathbf{R}+\mathbf{x})
  = \chi_{i_1i_2...i_n}^{(n-1)}(\mathbf{x})$,
$\Gamma_{i_1i_2...i_n}^{(n-1)}(\mathbf{R}+\mathbf{x})
  = \Gamma_{i_1i_2...i_n}^{(n-1)}(\mathbf{x})$, etc.

Within the Dzyaloshinski or generalized Coulomb gauge where the 
scalar potential $\phi(\mathbf{x},t) \equiv 0$, we may express 
both, the electric field $\mathbf{E}$ and the magnetic field 
$\mathbf{B}$, through the vector potential $\mathbf{A}$ according 
to
\begin{eqnarray}
\label{ref7}
\mathbf{E}(\mathbf{x},t)
  & = & - \pa_t \mathbf{A}(\mathbf{x},t),\\
\mathbf{B}(\mathbf{x},t)
\label{B-A}
  & = &   \nabla \times \mathbf{A}(\mathbf{x},t).
\end{eqnarray}
For this system, the Hamiltonian
that yields Maxwell's equations in the absence
of external sources and currents reads
\begin{eqnarray}
H(\mathbf{A},\mathbf{\Pi})
  & = &
\frac{1}{2} \int d^3x  
  \left( \frac{1}{\epsilon_0 \epsilon(\mathbf{r})}
         D_i(\mathbf{x},t) D_i(\mathbf{x},t) 
         + \mu_0^{-1} B_i(\mathbf{x},t) B_i(\mathbf{x},t) 
  \right)\nonumber \\ 
  & & + \frac{1}{3}\int d^3x
        \Gamma_{ijk}^{(2)}(\mathbf{x}) 
        D_i(\mathbf{x},t) D_j(\mathbf{x},t)D_k(\mathbf{x},t)+\dots
\nonumber \\
 & = & 
H_2 + H_3 + \dots\ .   \label{ref5}
\end{eqnarray}
In the former expression, $\mathbf{\Pi} = -\mathbf{D}$ denotes the
canonical momentum associated with the vector potential $\mathbf{A}$
\cite{hm}. Consequently, the equal-time Poisson brackets are
\begin{eqnarray}
\left\lbrace 
    A_{i}(\mathbf{x},t), \Pi_j(\mathbf{x'},t)
\right\rbrace_{\mathbf{A},\mathbf{\Pi}} 
  & = & 
\int  d^3 x''
\left( \frac{\delta A_{i}(\mathbf{x},t) }{  \delta A_{k}(\mathbf{x''},t)}
       \frac{\delta \Pi_j(\mathbf{x'},t)}{\delta \Pi_{k}(\mathbf{x''},t)}
       -
       \frac{\delta A_{i}(\mathbf{x},t) }{\delta \Pi_{k} (\mathbf{x''},t)}
       \frac{\delta \Pi_j(\mathbf{x'},t)}{   \delta A_{k}(\mathbf{x''},t)}
\right), \nonumber \\
& = & \delta_{ij} \delta(\mathbf{x}-\mathbf{x}'),
\end{eqnarray}
where ${\delta}/{\delta A_{i}(\mathbf{x},t)}$ and
      ${\delta}/{\delta \Pi_{i}(\mathbf{x},t)}$
denote functional derivatives (see Ref. \cite{morrison} 
for a definition and properties). The -- on first sight 
somewhat superfluous -- indices $\mathbf{A}$ and $\mathbf{\Pi}$ 
on the left-hand side allow us to keep track of the fields, once a 
transformation is applied to the Hamiltonian (\ref{ref5}).
To decompose this Hamiltonian into symmetry-adapted basis
functions (modes), it is useful to consider the solutions 
of the linearized equations of motion for periodic systems.

\subsection{Bloch function decomposition}
In the linear limit,  
$\mathbf{D}(\mathbf{x},t) =
  \epsilon_0\epsilon(\mathbf{x})\mathbf{E}(\mathbf{x},t)$, 
and the Maxwell equations reduce to two decoupled partial 
differential equations for the fields $\mathbf{A}(\mathbf{x},t)$ 
and $\mathbf{D}(\mathbf{x},t)$. Since we are interested in
stationary solutions (modes), we employ the time-harmonic ansatz 
$ \mathbf{A}(\mathbf{x},t) = \mathbf{A}_{n\mathbf{k}}(\mathbf{x})
                             \exp(-i\omega_{n\mathbf{k}}t)$, \ 
$ \mathbf{D}(\mathbf{x},t) = \mathbf{D}_{n\mathbf{k}}(\mathbf{x})
                             \exp(-i\omega_{n\mathbf{k}}t)$,
and obtain from the Maxwell equations
\begin{eqnarray}
\frac{1}{\epsilon(\mathbf{x})} \nabla \times
\left( \nabla \times \mathbf{A}_{n\mathbf{k}}(\mathbf{x})
\right)
  & = & 
\frac{\omega_{n\mathbf{k}}^2}{c^2} \mathbf{A}_{n\mathbf{k}}(\mathbf{x}), 
\nonumber \\
\nabla \times 
\left( \nabla \times 
       \left(
         \frac{\mathbf{D}_{n\mathbf{k}}(\mathbf{x})}{\epsilon(\mathbf{x})}
       \right)
\right) 
  & = &
\frac{\omega_{n\mathbf{k}}^2}{c^2} \mathbf{D}_{n\mathbf{k}}(\mathbf{x}).
\label{ref6}
\end{eqnarray}
In the above ansatz, we already made use of the Floquet-Bloch 
theorem that -- in the case of lattice periodic functions 
$\epsilon(\mathbf{x})$ -- applies to Eqs. (\ref{ref6}) and, 
therefore, introduced the Bloch functions 
$\mathbf{A}_{n\mathbf{k}}(\mathbf{x})$ 
and 
$\mathbf{D}_{n\mathbf{k}}(\mathbf{x})$.
These Bloch functions are labeled through the discrete band index 
$n>0$ and the continuous wave vector $\mathbf{k} \in \Omega_b$
that lies in the first Brillouin zone $\Omega_b$.
This labeling is commonly referred to as the reduced zone scheme
\cite{Ashcroft}.
In general, the Bloch functions and their associated eigenvalues
$\omega_{n\mathbf{k}}$ have to be determined numerically through
one of the well-documented techniques 
\cite{PhysRep}.
For future reference, we would like to note that the Bloch 
functions have the forms of modulated plane waves 
\begin{eqnarray}
\mathbf{D}_{n\mathbf{k}}(\mathbf{x})
  & = &
\mathbf{\tilde{d}}_{n\mathbf{k}}(\mathbf{x}) 
       \exp(i\mathbf{k}\mathbf{x}),
\label{ref11}\\
\mathbf{A}_{n\mathbf{k}}(\mathbf{x})
  & = &
\mathbf{\tilde{a}}_{n\mathbf{k}}(\mathbf{x})
\exp(i\mathbf{k}\mathbf{x}), 
\label{ref12}
\end{eqnarray}
where 
$\mathbf{\tilde{a}}_{n\mathbf{k}}(\mathbf{x})$ 
and 
$\mathbf{\tilde{d}}_{n\mathbf{k}}(\mathbf{x})$ 
are lattice-periodic functions. For convenience, the Bloch
functions $\mathbf{D}_{n\mathbf{k}},\ \mathbf{A}_{n\mathbf{k}}$
and the eigenvalues are chosen to be periodic in reciprocal
lattice vectors $\mathbf{G}$, i.e.
$\mathbf{D}_{n(\mathbf{k}+\mathbf{G})}=\mathbf{D}_{n\mathbf{k}}$
for all reciprocal lattice vectors $\mathbf{G}$ \cite{lax}.
It is also useful to note that the time-reversal symmetry of our
problem implies $\omega_{n\mathbf{k}} = \omega_{n -\mathbf{k}}$ 
which, in turn, implies that together with the solutions
$(\mathbf{D}_{n\mathbf{k}}, \nabla\times\mathbf{A}_{n\mathbf{k}})$, 
the corresponding pair 
$(\mathbf{D}_{n\mathbf{-k}}^{\star},-\nabla\times\mathbf{A}_{n-\mathbf{k}}^{\star})$ 
also fulfills the Maxwell equations for the same frequency.
Therefore, we choose 
$\mathbf{D}_{n\mathbf{k}}=\mathbf{D}_{n-\mathbf{k}}^{\star}$
which guarantees that the displacement field  
$\mathbf{D}(\mathbf{x},t)$ is real. The Bloch 
functions constitute a complete orthonormal set of basis 
functions. Since Eq. (\ref{ref7}) implies 
\begin{equation}
\label{ref8}
\mathbf{D}_{n\mathbf{k}}(\mathbf{x}) 
   = 
i \epsilon_0 \epsilon(\mathbf{x}) \omega_{n\mathbf{k}}
\mathbf{A}_{n\mathbf{k}}(\mathbf{x}),
\end{equation}
we find the the orthogonality relations
\cite{bvjs}
\begin{eqnarray}
\int d^3 x\ \epsilon_0 \epsilon(\mathbf{x})\ \omega_{n\mathbf{k}}\
\mathbf{A}_{n\mathbf{k}}(\mathbf{x})
\mathbf{A}_{n'\mathbf{k'}}^{\star} (\mathbf{x})
  & = & 
\frac{1}{2} \delta_{nn'}\delta({\mathbf{k}-\mathbf{k'}}),\\
\label{ref9}
\int d^3 x \left(
               \nabla\times \mathbf{A}_{n\mathbf{k}}(\mathbf{x})
           \right) \cdot
            \left(
               \nabla \times \mathbf{A}_{n'\mathbf{k'}}^{\star}(\mathbf{x})
            \right) 
  & = &  
\frac{\mu_0 \omega_{n\mathbf{k}}}{2} 
\delta_{nn'}\delta({\mathbf{k}-\mathbf{k'}}),
\end{eqnarray}
for $\mathbf{k},\mathbf{k}'\ \in\ \Omega_B$ if $n\neq0$.
For $n=0,\ \mathbf{k}=\mathbf{0}$, we have 
${\omega}_{n\mathbf{k}}=0$ and the normalization has to be 
performed separately. This issue will be addressed below.

For the subsequent formal developments, we find it more 
transparent to represent the Bloch functions in the so-called 
extended zone scheme 
\cite{Ashcroft}, 
where the wave vector $\mathbf{k}$ varies over the entire 
reciprocal space. Formally, this is facilitated by defining 
the band index to be a function of the wave vector, i.e., 
to introduce
\begin{equation}
\bar{\mathbf{A}}_{\mathbf{k}}(\mathbf{x})
  = 
\mathbf{A}_{n(\mathbf{k})\mathbf{k}}(\mathbf{x}),
\end{equation} 
and an analogous definition for $\bar{\mathbf{D}}_{\mathbf{k}}$
and $\bar{\omega}_{\mathbf{k}}$. Whether such a mapping does or does not
exist depends on the spectrum for the selected system, in particular
on the occurrence of degeneracies. 
For one-dimensional PCs and wave propagation along the axis of 
periodicity, wave vector and Bloch functions reduce to scalars. 
As a result, the unfolding of the band structure is realized 
through $n(k):=\lfloor {|k|d}/{\pi} \rfloor \ \in \mathbb{N}$,
where $\lfloor \cdot \rfloor$ denotes the Gauss bracket
defined via $0<|k|- (\pi/d) n(k)\le (\pi/d)$
($d$ denotes the lattice constant of the PC).
In two- or three-dimensional systems, this transformation 
between reduced and extended zone scheme is not straightforward 
but can, in principle, be carried out numerically. 
Therefore, we in the following restrict ourselves
to the case of strictly one-dimensional PCs and assume
that the system is periodic in the $x$ direction. 
To simplify the notation, we assume in the following an isotropic
nonlinear response of the constituent materials so that
the relevant fields ($\mathbf{A}$, $\mathbf{D}$, and 
$\mathbf{E}$) are all polarized along a certain transverse
direction which we define as the $y$ direction. As a result,
we may suppress the vectorial character of all fields.
In such a one-dimensional PCs, Eqs. (\ref{ref8}) - (\ref{ref9}) 
read as 
\begin{eqnarray}
\bar{D}_{k}(x)
 & = & 
i \epsilon_0 \epsilon(x)
  \bar{\omega}_{k} \bar{A}_{k}(x), \\
\int d  x\  
  \epsilon_0 \epsilon(x)\ \bar{\omega}_{k}
  \bar{A}_{k}(x) \bar{A}_{k'}^{\star} (x)
 & = & 
\f{1}{2} \, \delta({k-k'}),\\
\int d x\ 
  \pa_x \bar{A}_{k}(x)\ \pa_x \bar{A}_{k'}^{\star}(x) 
 & = &
\f{\mu_0 \bar{\omega}_{k}}{2} \, \delta({k-k'}),
\end{eqnarray}
for $k,k'\ \in\ \mathbb{R}\backslash\lbrace 0 \rbrace$.

We now return to the above-mentioned normalization problem 
related to $\bar{\omega}_{k=0}=0$ for which the eigenvalue
equation reads as 
\begin{equation}
\pa_{xx} \bar{D}_0(x) = 0.
\end{equation}
Owing to the Bloch-Floquet condition and our choice of 
$\bar{D}_k=\bar{D}^{\star}_{-k}$, we obtain 
$\bar{D}_0(x)=a \in \mathbb{R}$. On the other hand,
$\bar{D}_k$ must be a continuous function of $k$ so 
that the scaling $\bar{D}_k\propto \sqrt{\bar{\omega}_k}$
requires us to define
\begin{eqnarray}
\bar{D}_{k=0}
  & = &
0,\\ 
\lim_{\mu\rightarrow 0}
\bar{D}_{\mu}
  & \propto &
\lim_{\mu\rightarrow 0} \sqrt{\mu}.
\end{eqnarray}
Therefore, an absolute normalization in the sense of the 
orthogonality relations is impossible, yet our definition 
is justified from the physical point of view
\cite{ford1989}.
In fact, physically the mode $\bar{\omega}_{k=0}=0$ corresponds 
to a simple translation of our infinite PC and this clearly
requires an infinite amount of energy. Consequently, this mode
resists an absolute normalization. 
The above definition ensures that this is the case. Clearly,
this discussion may be directly transferred to the case of
two- and three-dimensional PCs.
Finally, we expand the physical fields $D(x,t)$ and $A(x,t)$
of the nonlinear system into the orthogonal basis of the 
linearized problem
\begin{eqnarray}
D(x,t)
  & = &
\int_{\mathbb{R}} d k
   \bar{D}_{k}(x) \bar{a}_{k}(t)\  +\ c.c.,\\
A(x,t)
  & = & 
\int_{\mathbb{R}} d k 
\bar{A}_{k}(x)\bar{a}_{k}(t)\  + \ c.c.,
\end{eqnarray}
where the time-dependent amplitudes 
$\bar{a}_{k}(t)$ and $\bar{a}_{k}^{\star}(t)$ 
are the weights associated with each Bloch mode. 
In terms of the physical fields, the new set of variables 
($k\neq 0$) are expressed as
\begin{equation}
\label{ref10}
\bar{a}_{k}(t)
   =   
\int d x \left( \f{1}{\bar{\omega}_{k}}
         \frac{\bar{D}_{k}^{\star}(x) D(x,t)}{ \epsilon_0 \epsilon(x)} 
       + \epsilon_0 \epsilon(x)\ \bar{\omega}_{k}\ 
         \bar{A}_{k}^{\star}(x)A(x,t)\right).
\end{equation}
In the derivation of Eq. (\ref{ref10}), the identity 
$(\bar{D}_{k}, \pa_x \bar{A}_{k})
 =
 (\bar{D}^{\star}_{-k}, -\pa_x\bar{A}^{\star}_{-k})$
has been employed and it can be verified
that Eq. (\ref{ref10}) represents a canonical transformation.
Consequently, for arbitrary functions $F,G$ in phase space,
\begin{equation}
\lbrace F,G \rbrace_{D,A} 
  = 
\lbrace F,G \rbrace_{\bar{a},\bar{a}^{\star}}.
\end{equation}
holds. In terms of the new variables, the equal-time Poisson
brackets now have the explicit form
\begin{eqnarray}
 \lbrace F,G \rbrace_{\bar{a},\bar{a}^{\star}} 
  & = & 
i \int_{\mathbb{R}} d k 
\left(
   \frac{\delta F}{\delta \bar{a}_{k}}
   \frac{\delta G} {\delta \bar{a}_{k}^{\star}} 
 - \frac{\delta G}{\delta \bar{a}_{k}}
   \frac{\delta F}{\delta \bar{a}_{k}^{\star}} 
\right),
\end{eqnarray}
so that 
\begin{eqnarray}
\lbrace \bar{a}_{k}, \bar{a}_{k'}^{\star} \rbrace_{\bar{a},\bar{a}^{\star}}
  & = & 
i  \delta(k-k'), \\
\lbrace \bar{a}_{k}, \bar{a}_{k'} \rbrace_{\bar{a},\bar{a}^{\star}} 
  & = & 0,
\end{eqnarray}
is immediate.
 
\subsection{Hamiltonian in the new variables \label{ref15}}
The results of the previous section may now be utilized to
rewrite the original Hamiltonian (\ref{ref5}) in terms of the
new field variables $ \bar{a}_{k} $ and $\bar{a}_{k}^{\star}$.
The linear part, $H_2$, is a quadratic form
\begin{equation}
H_2 
  = 
\int  d k\ \bar{\omega}_{k} \bar{a}_{k} \bar{a}^{\star}_{k},
\end{equation}
while the lowest-order nonlinear term, $H_3$, couples three
Bloch modes and reads as
\begin{eqnarray}
H_3
  & = & 
\ \int d k_{123} 
   V_{k_1k_2k_3}^{(1)} 
   \bar{a}_{ k_1} \bar{a}_{ k_2} \bar{a}_{ k_3} 
   \sum_{G}\delta(G-k_1-k_2 - k_3) \nonumber \\
  &   & 
+ \int d k_{123} 
   V_{k_1k_2k_3}^{ (2)} \bar{a}^{\star}_{ k_1} 
   \bar{a}_{ k_2} \bar{a}_{ k_3} 
   \sum_{G}\delta(G+ k_1- k_2 - k_3) \nonumber \\ 
  &   &
+\ c.c. \label{transam} \\
&=& H_3^{(1)}\ +\ H_3^{(2)}\ +\ c.c. \nonumber
\end{eqnarray}
In the former expression, the integration is carried out over all three 
momenta, i.e., $dk_{123} = dk_1 dk_2 dk_3$ and the sum over the 
entire set of reciprocal lattice vectors 
$G\ \in\ \lbrace G\in\mathbb{R}:\ G = m (2\pi / d) \ , 
         \ m \in\ \mathbb{Z} \rbrace$
reflects the fact that, in the extended zone description, in
principle all bands are coupled. In the reduced zone scheme, 
this would include both the so-called Umklapp processes, i.e., 
when the sum of the coupled wave vectors leaves the Brillouin 
zone as well as the  coupling between different bands without 
Umklapp processes. Indeed, this complication of classifying
and keeping track of the various scattering processes in the 
reduced zone scheme has lead us to adopt the (perhaps less
familiar) extended zone scheme. The corresponding coupling 
coefficients $V_{k_1k_2k_3}^{(1)}$ and
$V_{k_1k_2k_3}^{(2)}$ are derived from the Bloch functions
as given in Eqs. (\ref{ref11}) and (\ref{ref12}) 
(transferred to the extended zone notation) together
with the identity
\begin{equation}
\sum_{R} \exp\left(-i kR\right)
 =
\frac{2\pi}{d} 
\sum_{G} \delta(G+k),
\end{equation}
that allows one to reduce the integration over all space into
an integration over the PC's unit cell $V_{\rm cell}$
\begin{eqnarray}
V_{k_1k_2k_3}^{(1)} 
  & = & 
\frac{8 \pi^3}{3d} \int_{V_{\rm cell}} dx\
  \Gamma^{(2)}(x) 
  \bar{D}_{k_1} (x) \bar{D}_{ k_2} (x) \bar{D}_{ k_3} (x) \ \ , 
\label{ref20}\\
V_{k_1k_2k_3}^{(2)} 
  & = & 
\frac{8 \pi^3}{3d} \mathcal{P}_{c} \int_{V_{\rm cell}} dx\ 
  \Gamma^{(2)}(x) 
  \bar{D}^{\star}_{ k_1} (x) \bar{D}_{ k_2} (x)  \bar{D}_{ k_3} (x).
\label{ref21}
\end{eqnarray}
In this expression, we have introduced the cyclic permutation 
operator $\mathcal{P}_{c}$ that acts on any function $F_{123}$
with indices $123$ according to 
$\mathcal{P}_{c} F_{123} = F_{123}+F_{213}+F_{321}$.
Physically, the terms associated with $V_{k_1k_2k_3}^{(1)}$ correspond 
to the simultaneous creation or annihilation of three waves, while 
the terms associated with $V_{k_1k_2k_3}^{(2)}$ describe wave mixing 
processes where a single wave is converted into two other waves.
Clearly, a process associated to the matrix element $V_{k_1k_2k_3}^{(1)}$ does
not conserve 
energy. 
In order to be efficient, a coupling mechanism to another 
physical system is required. Therefore, for conservative optical
systems these processes are strongly suppressed.

The explicit form of the higher-order nonlinear contributions
to the Hamiltonian, $H_j$, $j>3$, follow in strict analogy to 
the derivation of $H_3$. For instance, the next nonlinear term 
$H_4$ contains both energy-suppressed and energetically allowed
processes. Among all of those, the relevant part of the
Hamiltonian corresponding to the third-order nonlinear 
susceptibility is
\begin{equation}
H_4^{(3)}
  =  
\int d k_{1234} 
   W_{k_1k_2k_3k_4}^{ (3)} 
     \bar{a}_{k_1} \bar{a}_{k_2} \bar{a}^{\star}_{ k_3} \bar{a}^{\star}_{ k_4}
     \sum_{G}\delta(G+ k_1+k_2-k_3 - k_4),
\end{equation}
where processes in which two waves are scattered are described
via the coupling coefficient
\begin{equation}
W_{k_1k_2k_3k_4}^{(3)} 
  = 
\frac{\pi}{2 d}
 \mathcal{P}_{6}\int_{V_{cell}} d x\ \Gamma^{(3)}(x) 
 \bar{D}_{k_1} (x,t) \bar{D}_{k_2} (x,t)  
 \bar{D}^{\star}_{k_3} (x,t)\bar{D}^{\star}_{k_4} (x,t).
\end{equation}
Here, the permutation operator $\mathcal{P}_{6}$ acts on a 
function with indices $F_{1234}$ according to
$\mathcal{P}_{6}F_{1234}
 =
F_{1234}+F_{1324}+F_{3214}+F_{1432}+F_{4231}+F_{4321}$.

The equations of motion for the canonical variables are the
Hamilton equations
\begin{equation}
\dot{\bar{a}}_{ k}(t)
 =
\lbrace \bar{a}_{ k}(t) , H \rbrace_{\bar{a},\bar{a}^{\star}}.
\end{equation}
However, it is often preferable to work in a frame that
``rotates'' with the frequency of the free carrier wave, i.e.,
to employ the slow variables 
$\bar{\alpha}_k(t) = \bar{a}_k(t) \exp(-i\bar{\omega}_k t)$. In this case
the equations of motions read
\begin{eqnarray}
\pa_t \bar{\alpha}_k
  & = & 
  \ i\int dk_{23} 
       V^{(2)}_{k k_2 k_3} \bar{\alpha}_{k_2}
       \bar{\alpha}_{k_3} e^{i(\bar{\omega}_{k_2}+\bar{\omega}_{k_3}
      -\bar{\omega}_k)t}\sum_G \delta(G+k-k_2-k_3) \\
  &   & + \lbrace \bar{\alpha}_{
k}(t) , H-H_3^{(2)} \rbrace_{\bar{a},\bar{a}^{\star}}
\end{eqnarray}
The terms on the right-hand side of the equation of motion give an
oscillatory contribution to the mode $\bar{\alpha}_k$ unless 
certain resonance conditions are met. For the one-dimensional photonic
crystals, these conditions are analyzed and compared to a related problem in ideal
hydrodynamics in the following section.

\section{Three-wave interaction}
\label{ref2} 
In many problems of interest in optics, the nonlinearities that 
are involved in the system dynamics are quite small. This 
suggests that in the equations of motion the nonlinear interaction 
terms can be considered as a perturbation to the linear part.
The resonance conditions for nonlinear wave mixing 
processes are determined 
by the linear dispersion relation $\omega_k$. 
A profound example of such a resonant mixing process in optics 
is the second harmonic generation, where the frequency of an 
incident beam is doubled through a quadratic, i.e., 
$\chi^{(2)}$-nonlinear response of the system. The requirement 
for efficient conversion is the conservation of energy and the
momentum in the process, $\omega+\omega = 2\omega$ and 
$k+k = 2k$, respectively. 
These equations are examples of the more general class of
three-wave resonance conditions
\begin{eqnarray}
s_{1}\omega(k_1)+s_{2}\omega(k_2) +  s_{3}\omega(k_3) 
 & = &
0\\
s_{1}  k_1 +s_{2}  k_2  +  s_{3}  k_3, 
 & = & 
0 \label{ref13}
\end{eqnarray}
where $s_i=\pm 1 $. 
These conditions are the requirements for an efficient energy 
transfer between three waves (modes) of wave numbers $k_1,k_2,k_3$.
For PCs, the second condition must be replaced by the conservation 
law for crystal momentum, i.e., reciprocal lattice vectors can be 
added to any of the momenta. For the moment, we leave this point 
aside but will return to it at a later stage of the discussion.
 
A detailed discussion of selection rules for nonlinear PCs can
be found in Ref. \cite{bf}. We would like to note that in the 
case of non-negligible third-order response, i.e., $\chi^{(3)}$ nonlinearities, similar
conditions must hold for four-wave mixing processes. 
At first sight, it appears difficult to solve Eqs. (\ref{ref13}) 
even if the linear spectrum $\omega_k$ is known analytically such 
as in the case of surface waves in an ideal fluid.
On the other hand, for fixed $s_i$, Eqs. (\ref{ref13}) contains 
six variables but only three of them can be chosen independently, 
and the two equations remove one degree of freedom. In this 
reduced parameters space and for any given $\omega_{k_0}$, one 
can easily carry out the analysis numerically and visualize the
resonance conditions.

Upon combining the expressions in Eqs. (\ref{ref13}), we obtain
\begin{equation}
\omega(-s_1s_2k_2 -s_1s_3k_3) + s_1s_{2}\omega(k_2) 
                              + s_1s_{3}\omega(k_3)=0 \label{ref14}
\end{equation}
Equation (\ref{ref14}) is invariant under the simultaneous inversion of 
the signs of all $s_i$, so that only four out of eight cases 
have to be considered. Time reversal symmetry, i.e.,
$\omega(k) = \omega(-k) \ge 0$, and the fact that in one-dimensional
PCs the spectrum is nondegenerate, lead to further reduction. In
fact, a semi-positive spectrum excludes the case of three identical 
signs and the time-reversal symmetry maps two other cases onto each
other up to an exchange of indices. Consequently, in order to find
nontrivial solutions to Eq. (\ref{ref14}), we are left with only two 
distinct cases. These can readily be visualized by defining the two 
functions
\begin{eqnarray}
a_1(k_2,k_3)
  & = & 
\frac{\omega(k_2-k_3)-\omega(k_2)}{\omega(k_3)}+1 \\
a_2(k_2,k_3)
  & = &
\frac{\omega(k_2+k_3)-\omega(k_2)}{-\omega(k_3)}+1 \label{f2}
\end{eqnarray}
for $k_2\in\mathbb{R}, \ k_3 \in\mathbb{R}\setminus\lbrace k_0=0
\rbrace$, ($k_3 = k_0$ only yields a trivial resonance). 
With these definitions, the resonance conditions are fulfilled
if and only if $a_i(k_2,k_3)=0$.

\begin{figure}
\centering
\includegraphics[width=3.2in]{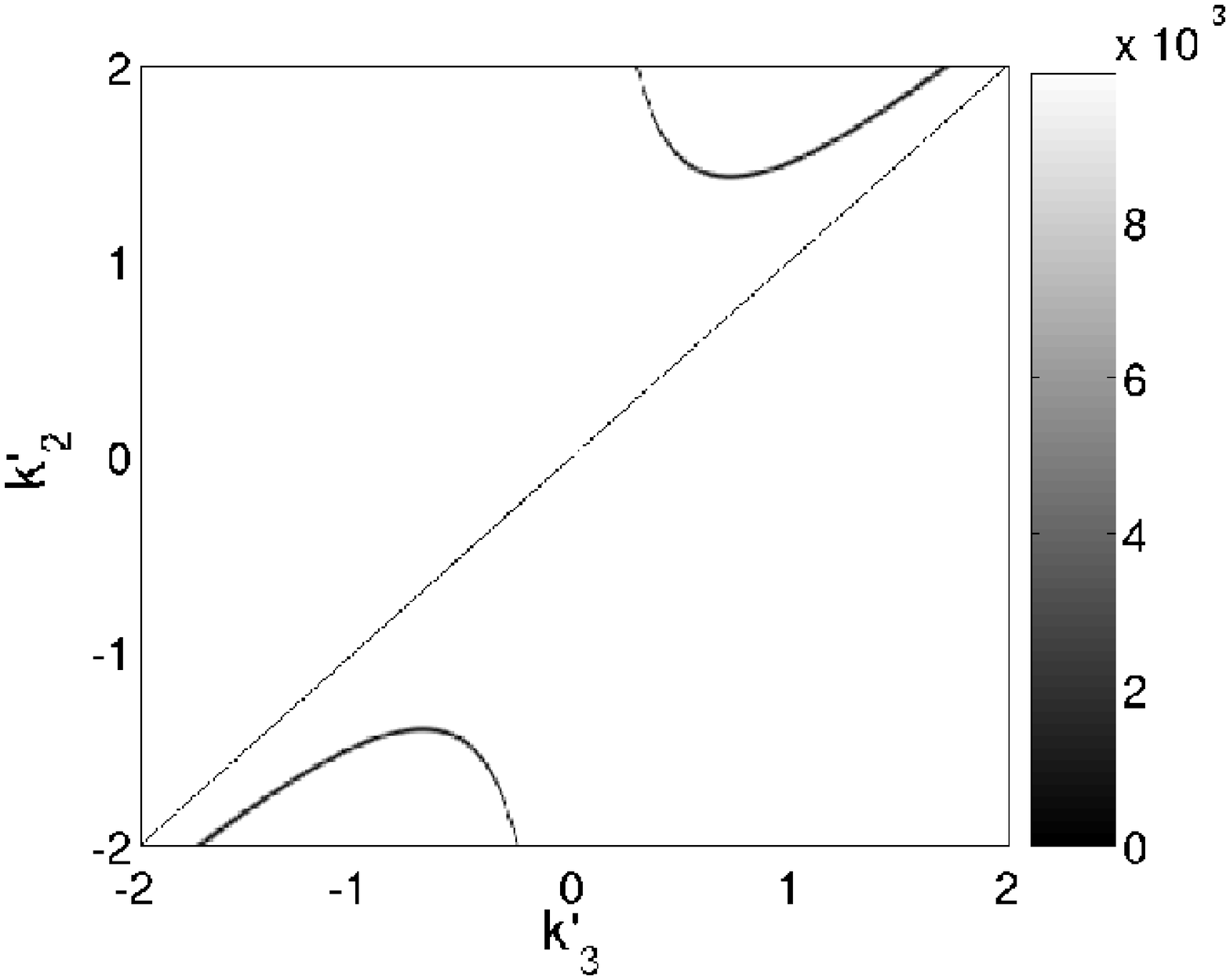}
\includegraphics[width=3.2in]{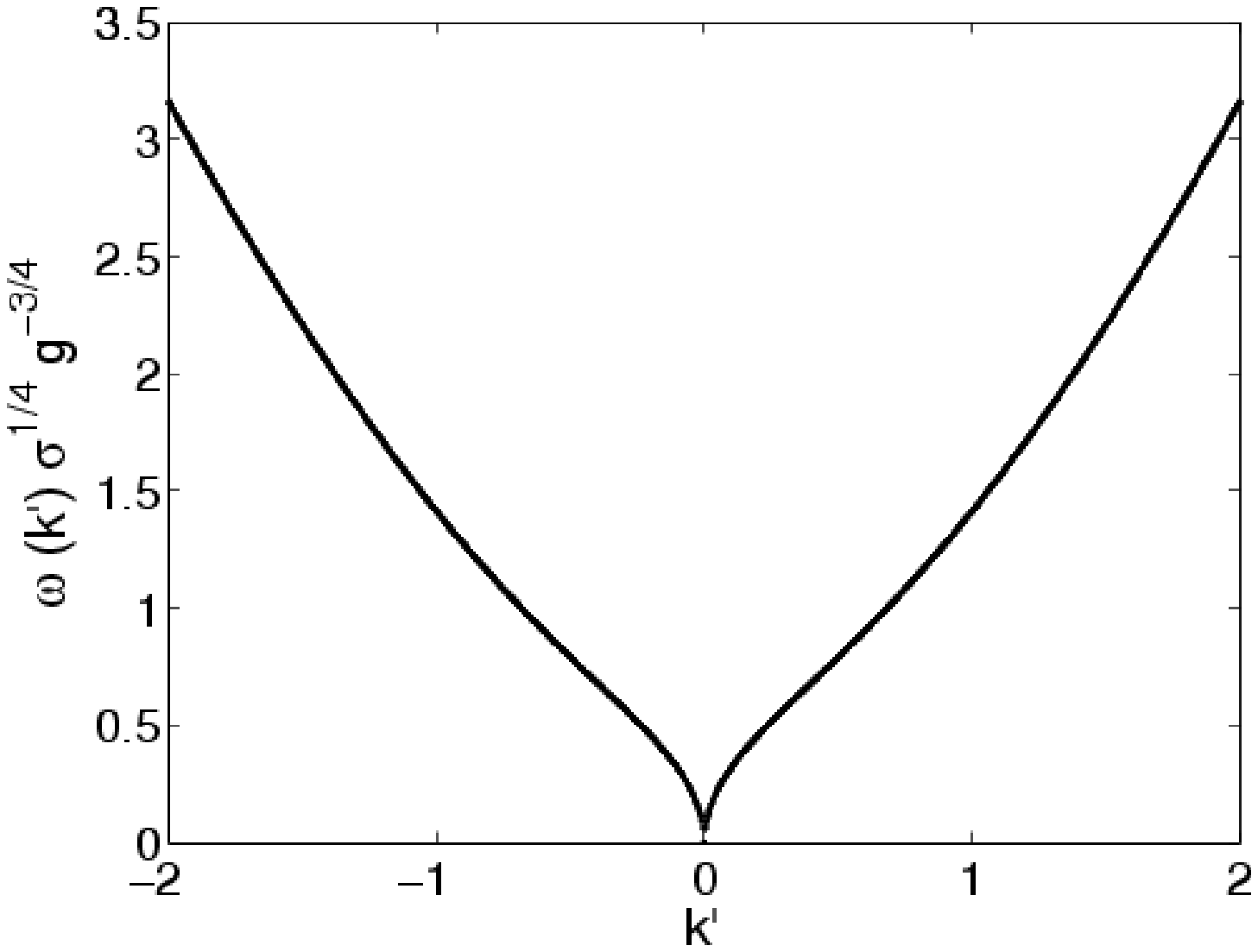}
\caption{Left panel : $|a_1(k'_2,k'_3)|$ for gravity-capillary waves;
         Right panel: Corresponding dispersion relation $\omega_{k'}$.
All variables are dimensionless, for notation and normalization see the text.
        } 
\label{fig1}
\end{figure}

The first spectrum $\omega_k$ for which we evaluate Eq. (\ref{f2}) 
is the dispersion relation for gravity-capillary waves on the surface
of an ideal fluid with infinite depth 
\begin{equation}
\omega_k = \sqrt{g|k|+\sigma|k|^3},
\end{equation}
where $g$ is the gravitational acceleration and $\sigma$ the
capillarity of the fluid. Introducing $k'=\sqrt{\sigma g^{-1}} k$
 (see also the right panel of Fig. \ref{fig1}), transfers the previous
expression into
\begin{eqnarray}
\omega_{k'} \sigma^{\frac{1}{4}}g^{-\frac{3}{4}}=\sqrt{|k'|+|k'|^3}
\end{eqnarray}
for evaluation in $a_1(k_2,k_3)$, $a_2(k_2,k_3)$. Notice that both $\omega_{k'} \sigma^{\frac{1}{4}}g^{-\frac{3}{4}}$ and $k'$ are dimensionless. It is, therefore,
possible to discuss the position of the resonances as a function of the
 gravitational and the capillary constant.
\begin{figure}
\centering
\includegraphics[width=3.2in]{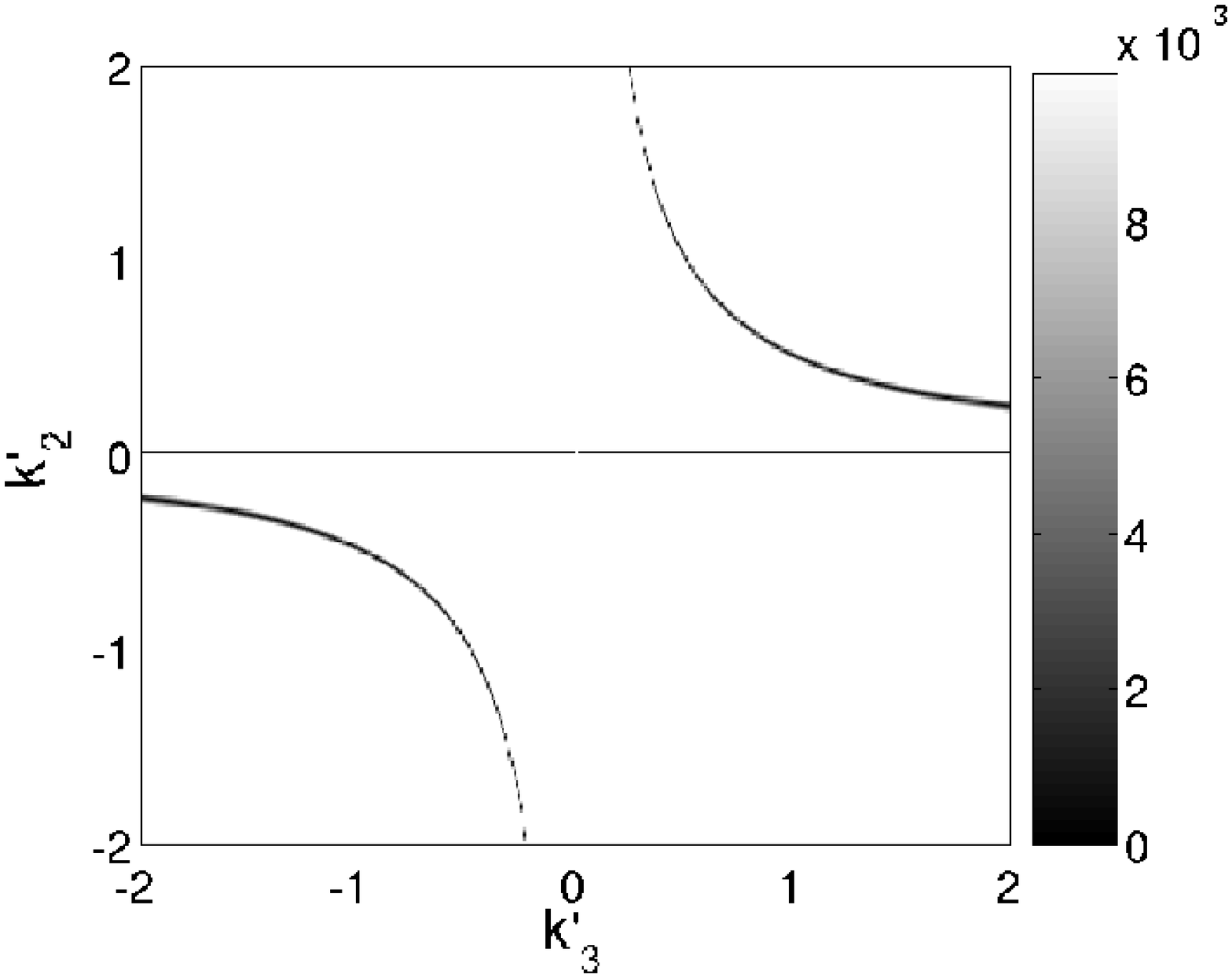}
\includegraphics[width=3.2in]{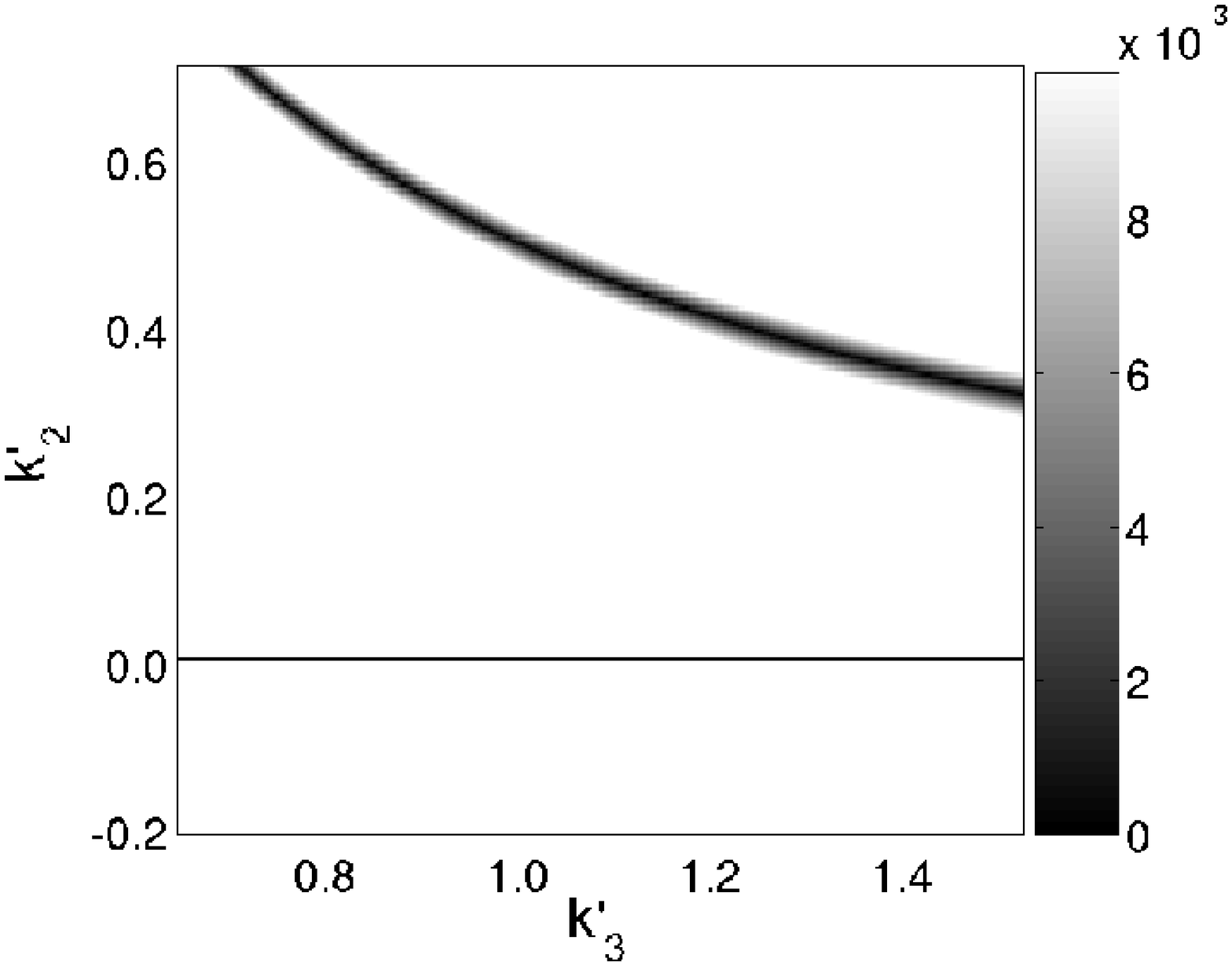}
\caption{Left panel  : $|a_2(k'_2,k'_3)|$ for gravity-capillary waves;
         Right panel : Close up showing the exact resonance near 
                       $k'_2=0$. 
All variables are dimensionless, for notation and normalization see the text.
        }
\label{fig2}
\end{figure}
The left panels of Figs. \ref{fig1} and \ref{fig2} reveal
the position of resonant or nearly resonant sets of wave vectors
$k'_2\in\mathbb{R},\ k'_3\in\mathbb{R}\setminus\lbrace 0\rbrace$,
i.e., sets for which 
$|a_i(k'_2,k'_3)|<0.01$. 
Black and grey regions correspond to 
$|a_i(k'_2,k'_3)|\ll 0.01$ 
or exactly zero, whereas white regions delineate wave vectors for 
which 
$|a_i(k'_2,k'_3)|\ge 0.01$. 
The symmetry with respect to the origin is a consequence of time 
reversal symmetry which implies that
$a_i(k'_2,k'_3)=a_i(-k'_2,-k'_3)$. 
Besides the trivial resonances 
$k'_2=k'_3$ for $a_1(k'_2,k'_3)$ 
(Fig. \ref{fig1}) and $k'_2=0$ for 
$a_1(k'_2,k'_3)$ (Fig. \ref{fig2}),
there are other sets of resonant and nearly resonant pairs of
wave vectors that are located on ribbons. It is instructive
to notice the influence of the capillary constant on the 
position of these ribbons by a scaling argument:
Selecting fixed $(k'_2,k'_3)$ in the set of nearly resonant
wave numbers, the location of the point moves in the direction of increasing $(k_2,k_3)$ with decreasing $\sigma$ such
that in the limit $\sigma\rightarrow 0$  these resonances lie in
the infinite for $(k_2,k_3)$.  For pure 
gravity waves ($\sigma = 0$), there are only trivial 
resonances when one of the wave vectors involves the dc component ($\omega_0=0$). The first 
nontrivial mixing processes occur at higher order (four-wave
processes).

\begin{figure}
\centering
\includegraphics[width=3.2in]{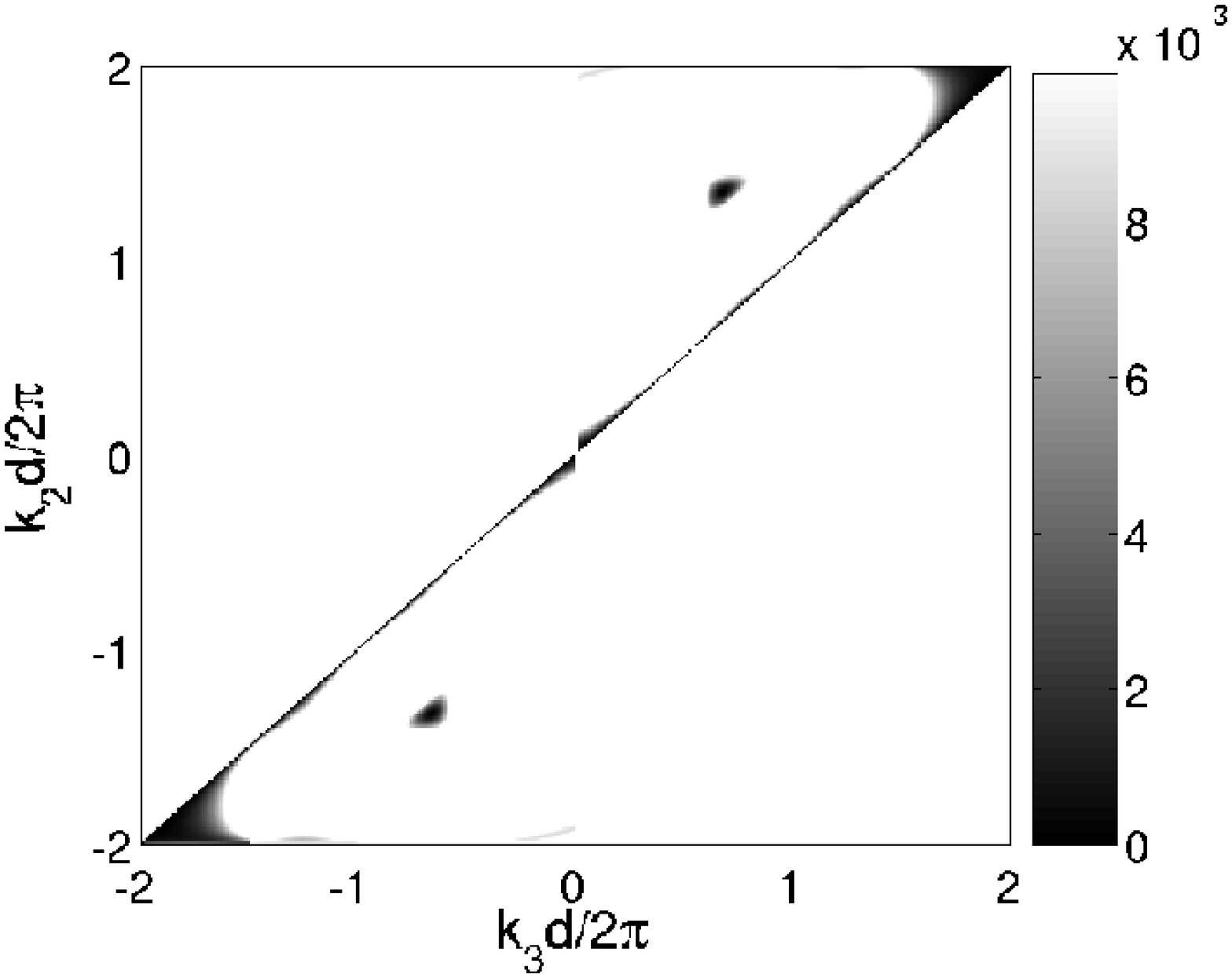}
\includegraphics[width=3.2in]{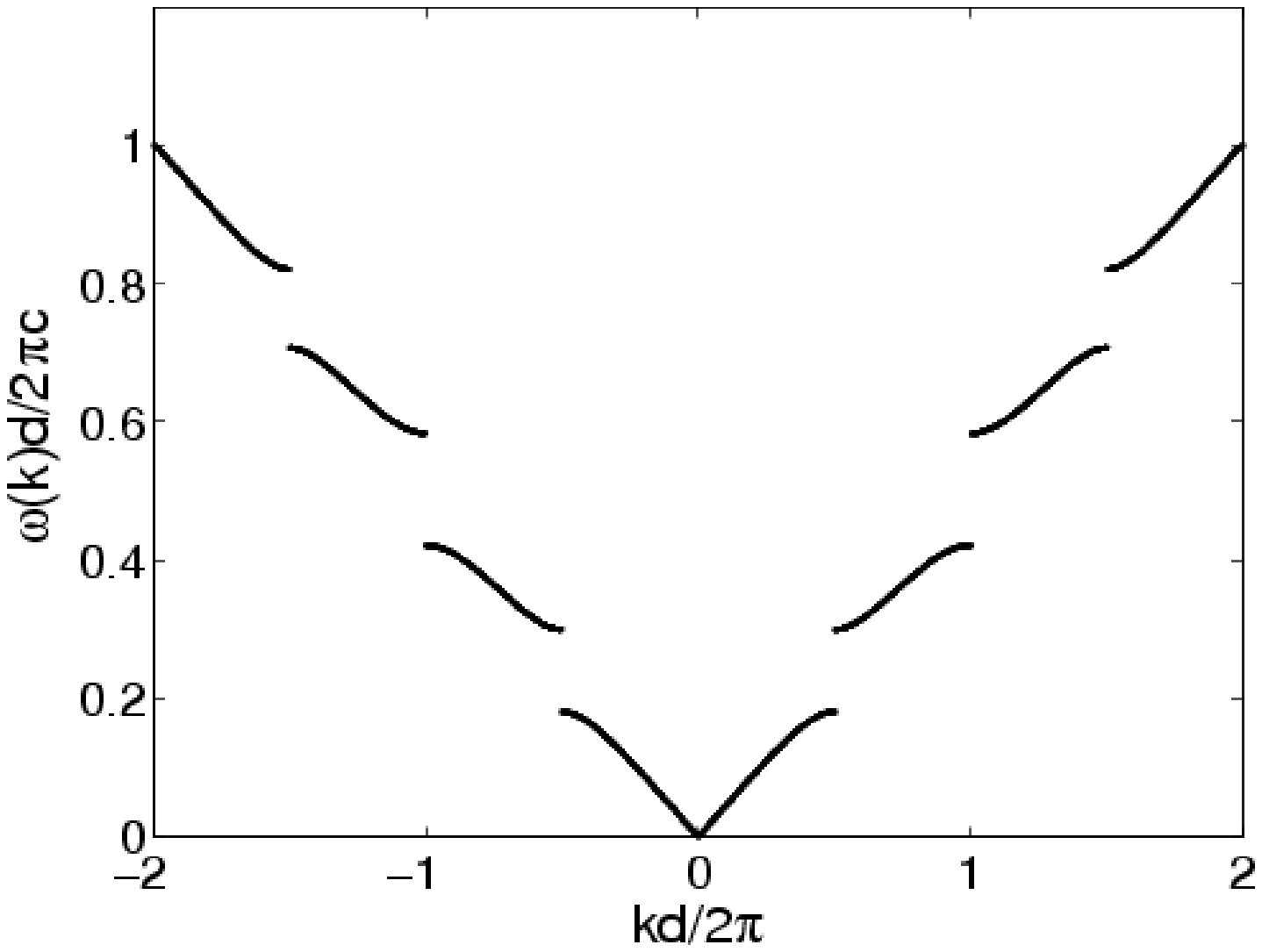}
\caption{Left panel : $|a_1(k_2,k_3)|$ for a PC with refractive index
                      contrast $\Delta n=2$; 
         Right panel: Corresponding photonic band structure $\omega_k$ 
                      in the extended zone scheme.
         } 
\label{fig3}
\end{figure}

Returning to three-wave mixing in PCs, the situation is very different from 
the previous case. In contrast to water waves, there is no explicit 
expression for $\omega_k$ and the discussion is necessarily 
qualitative.
In order to illustrate the situation, we assume that the PC consists
of alternating layers with equal thickness but different refractive 
indices, $n_A$ and $n_B$, respectively.
Then, the only parameter that alters the spectrum is the refractive
index contrast $\Delta n=n_A-n_B$. 
To calculate the photonic band structure, $\omega_k$, we use a 
plane-wave expansion method. Of course, any other method for obtaining photonic band structures could be used \cite{PhysRep}.
Except for the case of a homogeneous dielectric, the spectrum takes
the form of a band structure, as shown in right panel of 
Fig. \ref{fig3}.
\begin{figure}
\centering
\includegraphics[width=3.2in]{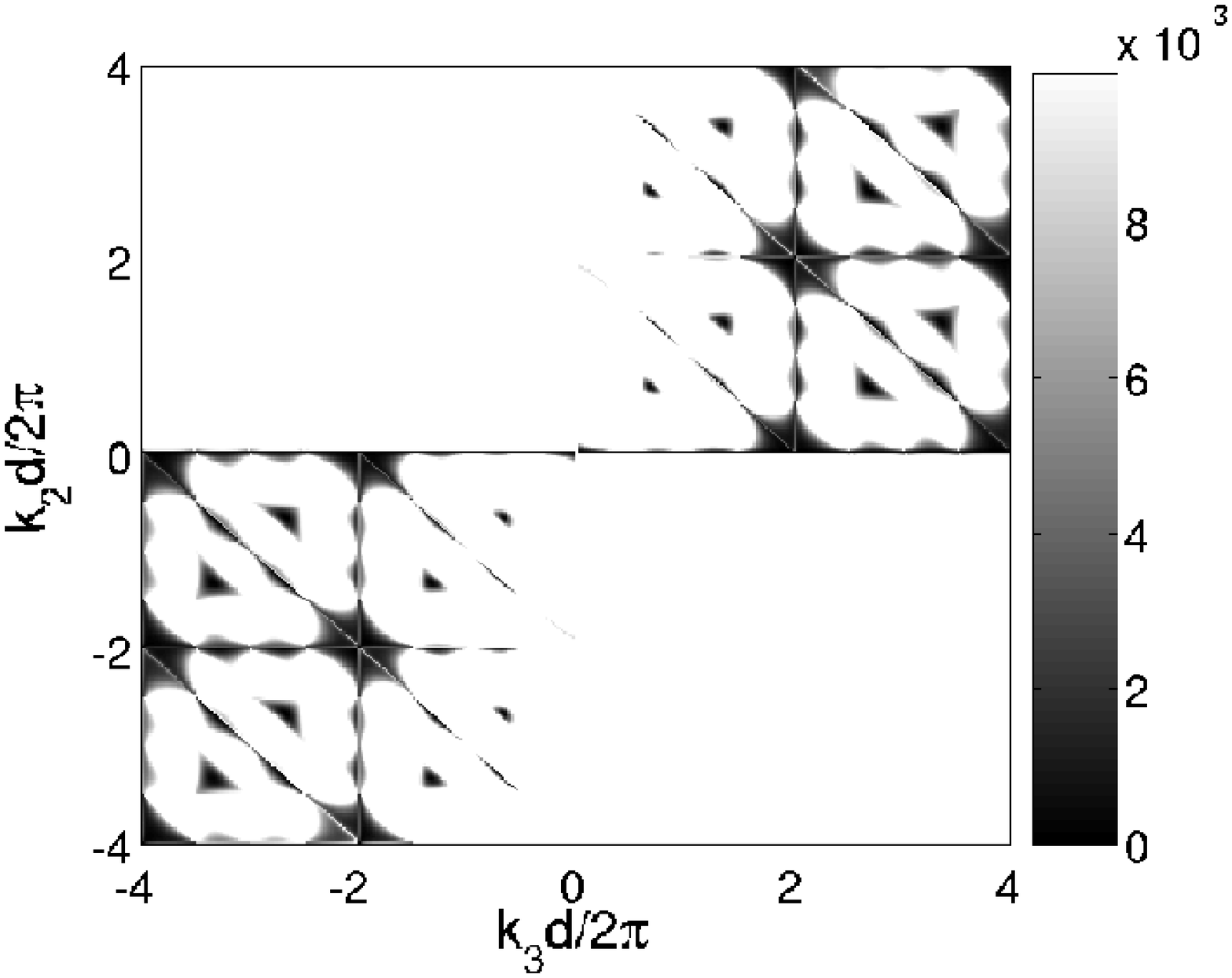}
\includegraphics[width=3.2in]{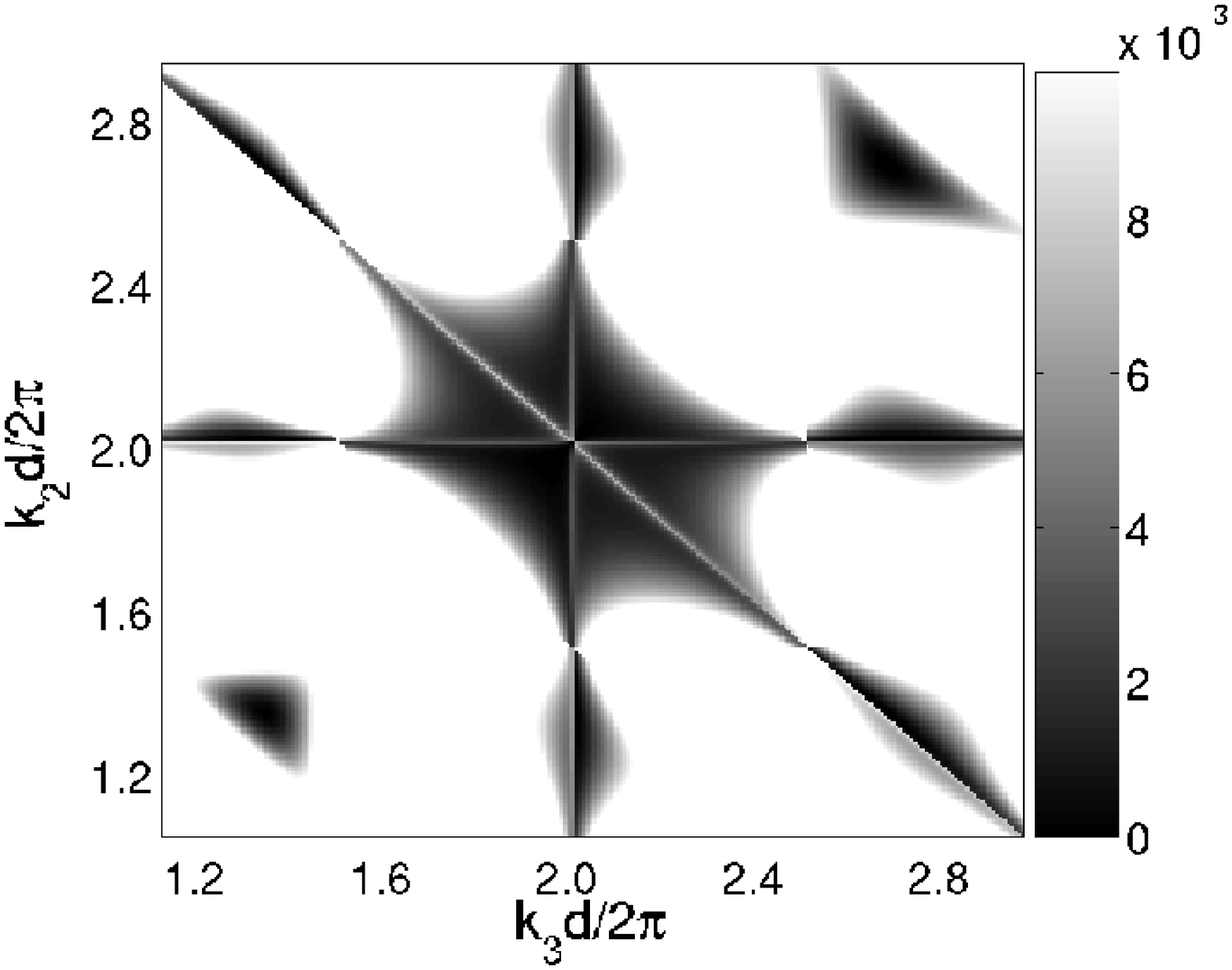}
\caption{Left panel : $|a_2(k_2,k_3)|$ for a PC with refractive index
                      contrast $\Delta n=2$; 
         Right panel: Close up of the upper right structure of 
                      the left panel.}
 \label{fig4}
\end{figure}
An example with an index contrast of 
$\Delta n = 2$  is shown in Figs. \ref{fig3} and \ref{fig4} and 
gives important qualitative information about the occurrence 
of resonances (dark regions) and quasi resonances (grey scale). 
In these figures, the wave vectors are multiplied with $d/2\pi$, consequently, the borders of successive Brillouin zones appear at 
half-integer values of the dimensionless variable. 
As in the case of water 
waves, the existence of $\omega_0=0$ leads to trivial resonances 
in the diagonal of Fig. \ref{fig3} and a horizontal line in Fig.
\ref{fig4}. In addition, there are isolated resonant and nearly
resonant subsets of wave vectors; their number and relative
position depends on the refractive index contrast. However, it is 
hard to quantify this dependence. In the extreme case of a 
homogeneous material, $\Delta n=0$, the resonance conditions can 
trivially be fulfilled. For instance, for $a_2(k_2,k_3)$ the two 
squares $k_2\ge 0,\ k_3>0$ and $k_2\le 0,\ k_3<0$ will become
identically zero. With increasing contrast, photonic band gaps form 
at the borders of the first Brillouin zone and the deviation from 
a linear spectrum is strongest in this region. This behavior is 
reflected by the pattern in the left panel of Fig. \ref{fig4}. 
As alluded to above, it is important to recall that in PCs only
crystal momentum is conserved, while the $a_i(k_2,k_3)$ in 
Figs. \ref{fig3} and \ref{fig4} have been calculated on the 
basis of momentum conservation. Therefore, for a complete 
interpretation all points in the plane $(k_2,k_3)$ have to be 
shifted through the reciprocal lattice vector 
$G=m 2\pi/d,\ m\in\mathbb{Z}$, both vertically and horizontally.
Consequently, there are more resonant zones than those shown 
in Figs. \ref{fig3} and \ref{fig4} that have to be considered.

Although at this point, we have not yet formally constructed the
normal-form Hamiltonian for the PC, we can already anticipate an 
important difficulty in its derivation. The resonance conditions,
Eqs. (\ref{ref14}), will appear in the denominators of the effective 
interaction coefficients. Whereas singularities resulting from 
trivial resonances are removable, the singular terms associated 
with nontrivial resonances are pertinent. Therefore, additional 
constraints on the distribution of wave vectors are required in 
order to ensure the applicability of the normal form transformation.
Since a wave packet in a nonlinear medium generally acquires new
wave vector components as time goes on, these constraints will
translate into constraints on the time scales for which our theory
will be applicable.  

\section{Reduction to normal form}
\label{ref3}
Normal form transformations represent a particularly useful tool
for the study of nonresonant wave interaction processes in 
Hamiltonian systems. In essence, they facilitate the reduction
of nonlinear dynamical systems to the simplest possible form
by eliminating inessential terms or degrees of freedom from
the original Hamiltonian. For physical systems with continuous degrees
of freedom, a good illustration of
this principle is the study of the aforementioned water wave 
problem which dates back to the early work of Zakharov
\cite{zakharov}. More recent highlights include the proof of the
nonintegrability of the water wave problem with the help of
normal form transformations by Craig and Worfolk 
\cite{Craig1,Craig2}.

In this section, we employ normal form transformations to
the case of nonresonant wave interaction in $\chi^{(2)}$ PCs.
The goal is derive an effective $\chi^{(3)}$ response of the
system that originates from a cascading of $\chi^{(2)}$ processes.
This analysis shows how the engineering of the PC structure
could, in principle, be used to tune the total $\chi^{(3)}$ response of the
system. 

In our analysis, we have to consider boundary 
conditions which lead to uncountably many modes. In addition, 
in PCs the resonant set of arguments cannot be unambiguously 
separated from the nonresonant ones since the nontrivial 
resonances have to be determined numerically and the influence 
of the refractive index contrast on their positions cannot 
be quantified (see Sec. \ref{ref2}). 
Similar to the cases of finite degrees of freedom with 
purely imaginary spectra of incommensurable eigenvalues
(depending on whether or not un-removable resonances remain,
the so-called Birkhoff of Birkhoff-Gustavson normal forms)
we utilize -- for our case of uncountably many degrees of 
freedom -- normal forms  as a formal tool to simplify the 
Hamiltonian and, hence, the equations of motion, but leave 
aside certain questions regarding the convergence of the 
transforms.

\subsection{Normal form transformation - Lie series}
In view of the discussion pertaining the efficiency of wave 
interaction processes in conservative systems (see Sec. 
\ref{ref15}), we assume in the following that the initial 
conditions, i.e. the initial distribution of modes 
$a_k(0), a_k^{\star}(0)$ in the PC with quadratic nonlinearity, 
are such that the resonance condition of lowest order that can 
be satisfied is a four-wave mixing process of the form
\begin{equation}
\omega_{k_1}+\omega_{k_2}-\omega_{k_3}-\omega_{k_4}=0.
\label{ref16}
\end{equation}
All other resonance conditions that may hold are assumed
to be of order 5 or higher. This four-wave resonance 
condition, Eq. (\ref{ref16}), is trivial in the sense that 
it can be fulfilled for any kind of dispersion relation and 
all possible initial conditions, even if the spectrum does 
not include the frequency $\omega=0$. Consequently, the 
system Hamiltonian 
$H=H_2+H_3+H_4+ \cdots$
comprises contributions in $H_3$ and $H_4$ that lead to 
 oscillatory source terms in the equation of 
motions for the time-dependent amplitudes 
$a_k(t), a_k^{\star}(t)$, as long as these
modes are bound away from the resonant subsets. 
To simplify the structure of these equations, we search for 
a coordinate transformation that eliminates all terms in the
Hamiltonian that correspond to resonance conditions which cannot 
be fulfilled. In constructing such a transformation, we follow
the ideas outlined in \cite{Craig1,Craig2,ckm}.

In order to split the Hamiltonian into  resonant and  
nonresonant parts, it is useful to introduce the linear 
mapping
\begin{equation}
ad_F(G) = \lbrace F,G\rbrace,
\end{equation}
where $\lbrace F,G\rbrace$ is the Poisson bracket of 
two functions $F,G$ in phase space. In quantum optics, the
operator $ad_{F}$ acting on any function $G$ is known as 
the Liouville operator. 
As an illustration, let us consider the Hamiltonian presented 
in Sec. \ref{ref1}. It can be easily verified that for 
$ H_3^{(1)}$ the following relation holds:
 $ \omega_{k_1}+\omega_{k_2}+\omega_{k_3}=0\ 
   \forall(k_1,k_2,k_3)\in\mathbb{R}^3\ 
  \Longrightarrow \ ad_{H_2}(H_3^{(1)})=0$.
More generally, $ad_{H_2}$ splits the Hamiltonian into resonant 
$H^{\rm res}_m$ and nonresonant $H^{\rm nr}_m$ terms of order 
$m$, where 
$H^{\rm res}_m\in ker(ad_{H_2})=\lbrace F_m:\ ad_{H_2}(F_m)=0\rbrace$.

Now, our goal is to eliminate all nonresonant terms from the 
Hamiltonian and to replace the initial system containing resonant 
and nonresonant terms with a simpler system. In particular, we 
want to find a canonical (symplectic) transformation  
$\bar{a}_k=F(\bar{b}_k)$, such that in the new variables $\bar{b}_k$ 
of the transformed Hamiltonian $H'(b)$ takes on the form
\begin{equation}
H'(b)=H_2(b)+H_4^{' \rm res}(b)+R''_5(b),
\end{equation}
where the residual $R''_5(b)$ contains only processes of fifth
or higher order in the amplitudes $b_k$.
Such reductions can be performed by means of the Lie series 
transformations, which in the nonlinear optics community received 
attention through the concept of the guiding center soliton in 
amplified fiber lines \cite{hk}. We define the Lie series 
$\exp(ad_K)$ induced by the generating functional $K$ as 
\begin{equation}
\exp(ad_K)(H)=\sum_{m=0}^{\infty} \frac{1}{m!} ad_K^{m}(H).
\end{equation}
Here, $K$ is a minimally third-order functional in the dynamic
variables. For the purposes of the present work, only a few
of the remarkable properties 
of these series will be used. Among them are \cite{ckm,groebner}:
\begin{eqnarray}
\exp(ad_K) \lbrace F,G \rbrace 
  & = & 
\lbrace \exp(ad_K) F , \exp(ad_K) G \rbrace\\
(\exp(ad_K))^{-1} 
  & = & 
\exp(ad_{-K}),\\
\exp(ad_K) F(u) 
  & = & 
F(\exp(ad_K)u),
\end{eqnarray}
where $F,G$ are functions in phase space and $u$ are the phase space variables. 
The first equality establishes the property of being a symplectic 
transformation for arbitrary generating functional $K$ and can be 
verified by using the Jacobi identity for the Poisson bracket. 
The second equality makes a connection to the inverse Lie series 
transformation. Finally, the last equation allows us to interchange 
the order of mapping. 

The explicit form of the polynomial $K$ can be determined by 
additional requirements on the normalization of the Hamiltonian. 
Writing $K=K_3+K_4$, where $K_3, K_4$, respectively, are  
-- yet to be determined -- functionals of order 3 and 4 
in the dynamical variables, and expanding the exponential yields
\begin{eqnarray}
\exp(ad_K)H
  & = &
\exp(ad_K)\left[H_2+H_3+H^{\rm res}_4+H^{\rm nr}_{4}+R_5\right] \nonumber\\
  & = &
H_2+H_3+H^{\rm res}_4+H^{\rm nr}_{4}+R_5+\lbrace K,H_2\rbrace \nonumber\\ 
  &   & 
+ \lbrace K,H_3\rbrace+ \f{1}{2}\lbrace K,\lbrace K,H_2\rbrace\rbrace+R'_5.
\end{eqnarray}
This suggests that the choice
\begin{eqnarray}
K_3 
   & = & 
ad_{H_2}^{-1}(H_3), \\
K_4 
   & = & 
ad_{H_2}^{-1}\left([H_4+\f{1}{2}ad_{H_3}(K_3)]_{\rm nr}\right)
\label{ref17}
\end{eqnarray}
realizes the desired simplification, i.e., we obtain
\begin{equation}
\exp(ad_K)H = H_2+H_4^{'res}+R''_5.
\end{equation}
In Eq. (\ref{ref17}),
 $[\cdot]_{\rm nr}$ indicates that we only consider the 
nonresonant terms contained within the bracketed expression.

Alternatively, we can apply a transformation from the original 
set of coordinates $\bar{a}_{k},\bar{a}^{\star}_{k}$ to a new set
$\bar{b}_{k},\bar{b}^{\star}_{k}$ according to
\begin{equation}
a_k=\exp(ad_{K(b)}) \, b_k,
\end{equation}
where $K$ is the functional defined in Eq. (\ref{ref17})
and we have -- and will continue to do so in the remainder of 
this paper -- dropped the bars over the coordinates. 
More explicitly, we choose
\begin{eqnarray}
K_3 
 & = & 
\int  dk_{123}\  
   A_{k_1 k_2 k_3}^{(1)}\ 
   b_{k_1} b_{k_2} b_{k_3} \sum_{G}\delta(G-k_1- k_2 -k_3) \\
  &  & 
  +  
\int d k_{123}\ 
   A_{k_1k_2k_3}^{(2)}\ 
   b^{\star}_{k_1} b_{k_2} b_{k_3} \sum_{G}\delta(G+k_1- k_2 - k_3)\ +\  c.c.,
\end{eqnarray}
as well as
\begin{eqnarray}
K_4
  & = & 
\int  dk_{1234}\  
   B_{k_1 k_2 k_3 k_4}^{(1)}\ 
   b_{k_1} b_{k_2} b_{k_3} b_{k_4} 
   \sum_{G}\delta(G-k_1- k_2 - k_3-k_4) \\
  &   & 
+ 
\int  dk_{1234}\  
   B_{k_1 k_2 k_3 k_4}^{(2)}\ 
   b^{\star}_{k_1} b_{k_2} b_{k_3} b_{k_4} 
   \sum_{G}\delta(G+k_1- k_2 - k_3-k_4)  \ 
   + \  c.c. \nonumber.
\end{eqnarray}
One can note that since 
$\left(\lbrace F,G \rbrace_{a,a^{\star}}\right)^{\star} 
  = 
\lbrace F^{\star},G^{\star} \rbrace_{a,a^{\star}}$, 
the requirement $K=K^{\star}$ follows immediately from the definition 
of the coordinate transformation.
The unknown coefficients $A^{(i)}$ in $K_3$ are determined from the
nonresonance condition, Eq. (\ref{ref17}), and are readily found 
to be
\begin{eqnarray}
A^{(1)}_{k_1k_2k_3}
  & = &
i \frac{V^{(1)}_{k_1k_2k_3}}{\omega_{k_1}+\omega_{k_2}+\omega_{k_3}},\\
A^{(2)}_{k_1 k_2k_3}
  & = & 
i \frac{V^{(2)}_{k_1k_2k_3}}{-\omega_{k_1}+\omega_{k_2}+\omega_{k_3}}.
\end{eqnarray}
Equation (\ref{ref17}) also fixes the coefficients $B^{(i)}$, 
which can be expressed in terms of $V^{(i)}$ and $A^{(i)}$, $i = 1, 2$. 
However, their explicit form is not required if we are only 
interested in terms at fourth order to which $K_4$ does not 
contribute. Finally, we note that the expressions
for $A^{(i)}$ become singular if the resonance condition is 
fulfilled.
 
In these new coordinates, the transformed Hamiltonian reads
\begin{eqnarray}
H'(b)
  & = & 
H_2(b)+H_4^{'res}(b)+R''_5(b) \nonumber\\
  & = & 
\int dk \, \omega_k b_k b_k^{\star} \nonumber \\
  &  & +  
\int d k_{1234} 
    S_{k_1k_2k_3k_4}
    b_{ k_1} {b}_{k_2} {b}^{\star}_{ k_3} {b}^{\star}_{ k_4} 
    \sum_{G}\delta(G+ k_1+k_2- k_3 - k_4)  \nonumber \\
  &   &
+ R''_5(b). 
\label{thm}
\end{eqnarray}
With the help of the corresponding Poisson brackets
\begin{eqnarray}
\lbrace b_k, b^{\star}_{k'}\rbrace 
  & = &
i \delta(k-k'), \\
\lbrace b_k, b_{k'}\rbrace
  & = &
0,
\end{eqnarray}
we finally obtain the transformed equations of motion  
\begin{eqnarray}
\frac{d}{dt} b_k 
  & = & 
i\omega_k b_k \nonumber  \\
  &  & 
+ i \int d k_{123} 
     (S_{k_1k_2k_3k}+S_{k_1k_2kk_3})b_{ k_1} {b}_{k_2} {b}^{\star}_{ k_3}  
     \sum_{G}\delta(G+ k_1+k_2- k_3 - k) \nonumber \\
  &  & 
+ \tilde{R}_4(b),
\label{ref18}
\end{eqnarray}
where $\tilde{R}_4(b)$ is a residual containing higher-order terms.

Equation (\ref{ref18}) represents the generalization of the 
Zakharov equation to the case of periodic media. This equation, 
which Zakharov reduced in his original article \cite{zakharov} to 
the Nonlinear Schr\"odinger equation (NLSE), has been the
subject of extensive studies for waves in homogeneous systems 
and has been tested experimentally for the case of surface water 
waves in channels \cite{skj}. 

\subsection{Effective fourth-order coefficient} 
Using the coefficients $A^{(i)}$, we now proceed to calculate 
$\lbrace K(b), H_3(b)\rbrace$ and its contribution to the 
transformed Hamiltonian, which contains the renormalized 
fourth-order matrix element.
Indeed, the effective fourth-order element $S_{k_1k_2k_3k_4}$ 
consists of the term originating from the third-order nonlinear 
susceptibility $\chi^{(3)}$ of the system and three additional 
terms $N^{(i)}$, $i = 1,2,3$ that arise from nonresonant 
$\chi^{(2)}$ processes
\begin{eqnarray}
S_{k_1k_2k_3k_4}
  & = &
W^{(3)}_{k_1k_2k_3k_4}
+\frac{1}{2}
  \left( N^{(1)}_{k_1k_2k_3k_4}
        +N^{(2)}_{k_1k_2k_3k_4}
        +N^{(3)}_{k_1k_2k_3k_4}\right) 
\label{ccoef}
\end{eqnarray}
where the explicit expressions for the $N^{(i)}$ are
\begin{eqnarray*}
 N^{(1)}_{k_1k_2k_3k_4}
  & = & 
-9 \sum_{G} \left(
\f{V_{(-G-k_1-k_2)k_1k_2}^{(1)}V_{(-G-k_1-k_2)k_3k_4}^{(1)\star}}{\omega_{(-G-k_1-k_2)}+\omega_{k_1}+\omega_{k_2}}
+\f{V_{(-G-k_3-k_4)k_3k_4}^{(1)\star}V_{(-G-k_3-k_4)k_1k_2}^{(1)}}{\omega_{(-G-k_3-k_4)}+\omega_{k_3}+\omega_{k_4}}
            \right),\\
N^{(2)}_{k_1k_2k_3k_4}
  & = &
\sum_{G} \left(
\f{V_{(G+k_1+k_2)k_1k_2}^{(2)}V_{(G+k_1+k_2)k_3k_4}^{(2)\star}}{-\omega_{(G+k_1+k_2)}+\omega_{k_1}+\omega_{k_2}}
+\f{V_{(G+k_3+k_4)k_3k_4}^{(2)\star}V_{(G+k_3+k_4)k_1k_2}^{(2)}}{-\omega_{(G+k_3+k_4)}+\omega_{k_3}+\omega_{k_4}}
         \right),\\
N^{(3)}_{k_1k_2k_3k_4}
  & = &
-4 \sum_{G} \left(
\f{V_{k_3(-G+k_3-k_1)k_1}^{(2)}V_{k_2(-G+k_3-k_1)k_4}^{(2)\star}}{-\omega_{k_3}+\omega_{(-G+k_3-k_1)}+\omega_{k_1}}
+\f{V_{k_1(-G+k_1-k_3)k_3}^{(2)\star}V_{k_4(-G+k_1-k_3)k_2}^{(2)}}{-\omega_{k_1}+\omega_{(-G+k_1-k_3)}+\omega_{k_3}}
            \right)\ .\\
\end{eqnarray*}
In the above expressions, the denominators of 
$N^{(i)}_{k_1k_2k_3k_4}$ 
contain the resonance conditions. In particular, frequencies which 
depend on the difference between two wave vectors appear. Note that 
$S_{k_1k_2k_3k_4}$ 
describes coupling of the four waves $b_{k_i}$, and therefore,  
trivial resonances appear in the denominators for any possible 
distribution of wave vectors in the wave packets $b_{k_i}$. 
As a consequence, 
$S_{k_1k_2k_3k_4}$ 
is ill defined, unless the singularity is removable. To analyze 
this issue in the next section, we further simplify the matrix 
elements by considering the simplest possible situation, i.e., the 
propagation of a monochromatic wave.

\subsection{Simple case - monochromatic wave}
In order to obtain further insight into the physics behind the coefficient, 
and to 
discuss the technical issue related to the appearance  of 
resonances, we consider the simple situation when initially 
only a single monochromatic wave is excited in the transformed 
space
\begin{equation}
b_{k}=B \, \delta(k-k_0), 
\end{equation}
where $B=const.\in\mathbb{C}$ and $k_0\neq 0$. In this case, 
the expression for the effective fourth-order element reduces 
to 
\begin{eqnarray}
 S_{k_0k_0k_0k_0} 
  & = & 
W^{(3)}_{k_0k_0k_0k_0} \nonumber\\ 
  &  &
-9 \sum_{G} 
    \f{\left|V_{(-G-2k_0)k_0k_0}^{(1)}\right|^2}{\omega_{G+2k_0}+2\omega_{k_0}}
+ \sum_{G}
    \f{\left|V_{(G+2k_0)k_0k_0}^{(2)}\right|^2}{-\omega_{G+2k_0}+2\omega_{k_0}}
-4 \sum_{G} 
\f{\left|V_{k_0(-G)k_0}^{(2)} \right|^2}{\omega_{G}}\ . 
\label{ref19}
\end{eqnarray}
At this point, the following comments are in order: 
\begin{itemize}
\item[(i)] 
    The terms arising from the nonresonant interaction processes 
    include a summation over all reciprocal lattice vectors $G$. 
    Since the coefficients 
    $V^{(i)}_{k_1k_2k_3}$ 
    and frequencies 
    $\omega_{k_i}$ 
    are evaluated in the extended zone scheme, a summation over 
    reciprocal lattice vectors $G$ is equivalent to a summation 
    over all band indices $n$. 
\item[(ii)] 
    For $G=0$, the last term is potentially undefined since 
    $\omega_0=0$. Fortunately, this singularity is removable. 
    Indeed, as discussed in Sec.~\ref{ref1}, the Bloch 
    functions are normalized as $D_k\propto\sqrt{\omega_k}$. 
    Therefore, we obtain
    \begin{equation}
    \lim_{\mu\rightarrow 0} 
    \frac{\left|V_{k_0 \mu (k_0+\mu)}^{(2)}\right|^2}{\omega_{\mu}}
      =  
    \lim_{\mu\rightarrow 0}
    \frac{\omega_{k_0} \omega_{\mu} \omega_{k_0+\mu}
    C(k_0,\mu)}{\omega_{\mu}}  
      = 
    \omega_{k_0}^2 C(k_0,0),
\end{equation}
    with some constant $C(k,0)$ for which 
    $0 \le C(k,0)<\infty\ \forall (k)\in\mathbb{R}$. 
\item[(iii)] 
     The denominator of the terms in the second line of Eq. (\ref{ref19}) 
     include the squared modulus of the initial three-wave matrix 
     element. From Eqs. (\ref{ref20}) and (\ref{ref21}) it is clear that
     the nonlinear susceptibility enters with a power of 2 into 
     the numerators.
     We would like to note that the resonance conditions appearing 
     in the denominators of $N_i$ correspond precisely to the 
     three-wave interactions that have been eliminated from the 
     Hamiltonian. As a result of this elimination, these processes
     give effective contributions to the interactions arising from 
     nonlinear coupling via $\chi^{(3)}$.
     Considering the contribution of $N^{(2)}_{k_0k_0k_0k_0}$, the 
     phase mismatch of second harmonic generation appears in the 
     denominator, whereas for $N^{(3)}_{k_0k_0k_0k_0}$, the sum of 
     frequencies accounts for the interaction of two waves with the 
     same frequency via the static field $\omega_0$ or via waves at 
     symmetry points $\omega_G$. In the field of nonlinear optics 
     of dispersive homogeneous media, effective higher-order nonlinear 
     contributions due to phase mismatch are generally referred to
     as cascaded processes. These effects are known to effectively 
     modify the nonlinear third-order response of homogeneous materials 
     with noncentrosymmetric nonlinear susceptibilities 
     \cite{bszg}. 
     The phase mismatch in these materials results from intrinsic 
     material related dispersion, which -- due to Kramers-Kronig
     relations -- always appears together with absorption. 
     Although some progress has recently been made on the 
     Hamiltonian description of the optical properties of nonlinear 
     dispersive materials 
     \cite{sw}, 
     the Lie series transformation
     does not directly apply to this class, because the detailed 
     knowledge of interaction coefficient relies on a modal description. 
     The key aspect about PCs which enable the formal construction of 
     a normal form is that nontrivial dispersion relations result from 
     the formation of a photonic band structure. Consequently, if
     the validity of the equation could be justified for finite times 
     with an appropriate error bound, the result suggest that it is 
     possible to achieve cascading without using strongly dispersive 
     and, therefore, lossy absorbing materials. In addition, the 
     judicious engineering of PCs would allow to tailor the 
     effective third-order nonlinear response of these systems
     through both the dispersion relation $\omega_k$ and the
     matrix elements $V^{(1)}_{k_1 k_2 k_3}$ and 
     $V^{(2)}_{k_1 k_2 k_3}$.
     Taken together, these two aspects enlarge the possibilities 
     for the realization of optical materials with tailored 
     nonlinear properties. 
\end{itemize}
Although some results on the validity for finite times of 
envelope equations that have been derived under nonresonance 
assumptions in the presence of resonances  are known \cite{gs05}, 
these turn out to be inapplicable to the present case, as outlined 
in the following section.
 
\subsection{Influence of resonances}
As a consequence of the strong asymptotic resonance 
\begin{equation}
r(k,m) = \omega (k) - \omega (k-m) - \omega (m) \to 0
\end{equation}
for $k,m \to 0$, the system truncated after the third order terms
cannot be expected to be well posed. In detail, we have 
$\omega_k \sim k$ plus some correction terms coming from the 
periodic coefficients. It is well known \cite{Eastham73} that 
the corrections decay asymptotically with the regularity of 
the associated coefficients. 
If the coefficients are $n$-times differentiable, then for suitably 
chosen
$k$ and $m$, we obtain
\begin{equation}
r(k,m) \leq C\left( |k|^{-n} + |k-m|^{-n} + |m|^{-n}\right).
\end{equation}
Since $r(k,m)$ appears in the denominator in the normal form 
transformation, there is a substantial loss of regularity 
associated with the normal form transformation. Thus, the 
less regular the coefficients, the better are the conditions 
for well posed truncated third-order systems. 

The same problems occur if the truncated third-order system 
is justified with some approximation result similar to that 
of Ref. \cite{BSTU06}. In this work, the NLSE has been justified 
in the above sense for semi linear wave equations with periodic 
coefficients. An additional difficulty occurs in our situation 
due to the quasilinearity of the problem that always leads 
to some loss of regularity in normal form transformations.

However, on physical grounds (see Sec. \ref{ref2}) we expect 
that for suitably chosen initial conditions, the dynamics of the 
system is dominated by nonresonant processes for not too long 
times. Quantifying this conjecture by methods of asymptotic analysis is 
a challenging mathematical problem that has not been addressed yet. 
\section{Conclusions \label{ref4}}
In this article, we have extended a recently developed Hamiltonian 
formalism to the analysis of nonresonant wave interaction in 
one-dimensional PCs with quadratic nonlinearity. 
In particular, we have shown that an appropriate normal form 
transformation leads to a generalization of Zakharov's equation 
for periodic media.
For the simplest case of monochromatic wave, the resulting 
effective coupling coefficient in the transformed equations 
of motion is the result of cascading second-order nonlinear processes. 
Moreover, we derived explicit expressions for the emerging effective 
nonlinear coupling terms. 
This may allow one to tailor the effective nonlinear response 
of the composite materials through judiciously engineering of its 
linear properties and could,  in turn, suggest possibilities 
for the realization of optical
materials with customized nonlinear properties. The approach presented here
can also be applied to nonresonant higher-order interaction 
processes. 
Owing to the linear dispersion relation in the long-wavelength 
limit certain mathematical difficulties arise that limit the 
applicability of the method to finite times. However, we would
like to emphasize that in experiments, various loss mechanisms 
or the finite sample size set maximal time scales so that this
limitation may be less severe. In addition, full numerical 
simulations can provide further insight into the range of
validity of our theory.
A systematic study of the Zakharov equation's validity represent 
a challenging mathematical task and further research is indicated. 

\textbf{Acknowledgements:}
JGH, LT, and KB acknowledge support from the DFG-Forschungszentrum 
Center for Functional Nanostructures (CFN) at the Universit\"at 
Karlsruhe within project No. A1.2. The research of LT and KB is further 
supported through the DFG-Priority Program No. SPP 1113 {\it Photonic 
Crystals} within project Bu 1107/6-1. LT also acknowledges support from USA CRDF Grant No. GEP2-2848-TB-06.
Finally, the authors would like to thank J. Niegemann, L. Shemer, 
and  H. Uecker for fruitful discussions, and S.N. Volkov and 
J.E. Sipe for helpful correspondence.

\bibliographystyle{apsrev}

\end{document}